\documentclass[%
 reprint,
 amsmath,amssymb,
 aps,
]{revtex4-1}

\usepackage{graphicx}
\usepackage{dcolumn}
\usepackage{bm}

\newcommand{\qq}{\qquad}
\newcommand{\ket}[1]{\left|{#1}\right\rangle}
\newcommand{\bra}[1]{\left\langle{#1}\right|}
\newcommand{\aver}[1]{\left\langle{#1}\right\rangle}

\usepackage{color}
\usepackage{hyperref}
\usepackage{caption} 
\usepackage{booktabs,siunitx}
\newcommand{\be}{\begin{equation}}
\newcommand{\ee}{\end{equation}}
\newcommand{\bea}{\begin{eqnarray}}
\newcommand{\eea}{\end{eqnarray}}
\usepackage{amsmath,amssymb}

\begin{document}

\title{Time-series and network analysis in quantum dynamics: \\
Comparison with classical dynamics}

\author{Pradip Laha}
 \altaffiliation[]{pradip@physics.iitm.ac.in}
\author{S. Lakshmibala}%
\affiliation{%
 Department of Physics, IIT Madras, Chennai 600036, India\\}%
\author{V. Balakrishnan}%
\affiliation{Department of Physics, IIT Madras, Chennai 600036, India\\}%

\date{\today}

\begin{abstract}
Time-series analysis and network analysis are now
 used extensively in diverse areas 
of science.  In this paper, we apply  
these techniques to  quantum dynamics in an 
optomechanical system: specifically, 
the long-time dynamics of  the mean photon number  in 
an archetypal  tripartite quantum system comprising   a single-mode  radiation field interacting with a two-level atom and an oscillating membrane. We also investigate a  classical system of interacting Duffing oscillators which effectively mimics several of the features of tripartite quantum-optical systems. In both cases, we examine the manner in which  the maximal Lyapunov exponent obtained from a detailed time-series analysis varies with changes in an appropriate tunable parameter of the system. Network analysis is employed in both the quantum and classical models to identify suitable network quantifiers which will reflect these variations with the system parameter. This is a 
novel  approach  towards (i) 
examining how a considerably smaller data set (the network) obtained from a long time series of dynamical variables captures important aspects  of the underlying dynamics, 
and (ii) identifying  the differences between classical and quantum dynamics. 
 
\begin{description}
\item[PACS numbers]
05.45.Tp; 42.50.-p; 05.45.-a
\end{description}
\end{abstract}

\pacs{Valid PACS appear here}
\keywords{
Cavity optomechanics, Duffing oscillator, time-series analysis, network analysis, recurrence plot, maximal Lyapunov exponent}
\maketitle


\section{\label{sec:level1} Introduction}
\noindent  The availability  of time-series data in diverse areas  such as  weather forecasting, climate  research and  medicine~\cite{zou,gao1,donges,marwan2,ramirez}  has facilitated  detailed investigations leading to the extraction of important results on the dynamics of a variety of systems. Several tools have been proposed in  time-series analysis  to assess the long-time behaviour of  complex dynamical systems. The methods used  involve the identification and estimation  of indicators of the nature of the underlying dynamics such as the  maximal Lyapunov exponent (MLE),  return maps, return-time distributions, recurrence plots,  and  so on.  

In recent years, the 
 analysis of  networks constructed from a long time series   has proved to be  another important tool that has contributed significantly to the understanding of  classical dynamics~\cite{newman_book,cohen,newman,boccaletti,kurths_phys_rep}. The problem of handling a large 
 data set is circumvented by reducing it  to a considerably smaller optimal set (the network),  particularly in the context of machine learning protocols~\cite{seth_lloyd,rebentrost}.  
 Different methods  
 have been employed 
 to convert the time series of a  
 classical dynamical variable into an equivalent network, 
each method  capturing specific features of the dynamics encoded in the time series~\cite{zhang,lacasa,nicolis,
  marwan1,yang,xu,donner_epj}. In this paper, we have  constructed  $\epsilon$-recurrence networks   to obtain  smaller data sets from the  time series of relevant observables of certain tripartite systems. We have carried out this investigation in the context of both quantum and classical dynamics. The network indicators that we 
 consider   are the average path length (APL), link density (LD) clustering coefficient (CC), transitivity, assortativity and degree distribution. The purpose of this study is three-fold: (a)~to examine  the manner in which these network  indicators vary  with changes in  specific system parameters; (b)~to  assess the extent to which these variations reflect those of  indicators  obtained from the full data set such as the MLE; (c)~to understand the differences in the behavior  of  network indicators 
 computed from  data sets pertaining, respectively, to quantum and classical systems.  

 In the quantum mechanical context, we have examined the time series data for the mean photon number of the 
 radiation field in a cavity optomechanical system 
 as well as the equivalent network. 
 The results obtained have been 
compared  with corresponding results 
reported in an earlier work~\cite{laha4} 
for another tripartite quantum system, namely, 
a three-level $\Lambda$-atom interacting with 
two radiation fields. (In what follows, we shall refer to this system as the tripartite $\Lambda$ system). The optomechanical model  involves the interaction between the optical field contained in a cavity with a two-level atom placed inside the cavity, and a mechanical oscillator attached to one of the cavity walls, which is capable of small oscillations.  The oscillations as also the atomic transitions  are governed by the radiation pressure.  The dynamics of  the quantum oscillator has been {\em  controlled}  by this method  in several contexts, 
such as the detection of gravitational waves~\cite{abramovici,braginsky}, high precision measurements of masses and the weak force~\cite{vitali2,geraci,lamoreaux},  quantum information  processing~\cite{stannigel}, cooling mechanical resonators very close to their quantum ground states~\cite{barzanjeh,wilson_rae,genes1,li}, and examining classical-quantum  transitions in mechanical systems~\cite{schwab,marshall}. Optomechanical systems have thus attracted  considerable attention   
both  theoretically as well as  experimentally 
(see also ~\cite{aspelmeyer,bowen} and 
references therein).   

Further, if the field-atom coupling is dependent on the field intensity, new phenomena occur. A special form of  the  intensity-dependent coupling (IDC) which is important from a group-theoretic point of view is given by  $f(N) = (1 + \kappa \,N)^{1/2}$ where $\kappa$ is the `intensity parameter'  and $N$ is the photon number operator~\cite{siva}. It has been shown in  earlier work~\cite{lahap} that, for this 
form of the IDC,  the dynamics of  the mean photon number $\aver{N}$ as well as  the 
entanglement properties depend sensitively on $\kappa$. These interesting features in the dynamics make this model a good candidate for time-series and network analysis. 

 The classical system we consider  here is a set of two coupled Duffing oscillators. The dynamical variable in this case is essentially the velocity  of one of the oscillators. As is well known, 
 the Duffing oscillator exhibits rich dynamical behaviour (see, for instance, ~\cite{alzar}), which makes it an ideal candidate for examining generic features of time series and networks,  so that inferences can be drawn 
 in a general setting. The Duffing equation 
 has been extensively used  
 to model  the behaviour of a wide spectrum of mechanical oscillators, electrical circuits, nonlinear pendulums, 
 aspects of hydrodynamics, and so on. Small 
 variations  in the system parameters can produce significant changes in the dynamics,  
  ranging from quasiperiodicity to chaos~\cite{argyris}. 
 
 The reason for focusing on the system of 
 Duffing oscillators for our purposes 
 is as follows. The phenomenon of 
 electromagnetically induced transparency (EIT) occurs under suitable conditions in quantum systems involving 
 an  atomic medium interacting with two laser fields  
 (see, for instance,~\cite{harris}). EIT basically refers to the appearance of a transparency window within the absorption spectrum of the atomic system.  This effect  has been observed in many experiments,  and several investigations  have been carried out using theoretical  models that explain the occurrence of EIT.  A  simple  quantum system exhibiting EIT is the tripartite $\Lambda$ system mentioned earlier. Of immediate interest to us is the fact that a classical analog  of EIT-like behavior has been demonstrated in 
 as simple a  system as two coupled harmonic oscillators subject to a harmonic driving force~\cite{alzar}. Inclusion of  a cubic nonlinearity and dissipation leads to more interesting  and physically more realistic 
 behavior, which can be effectively modelled by two coupled Duffing oscillators.  
 Motivated  by the  diversity of  its dynamics  and its 
 capability  to mimic certain types of quantum 
 phenomena  such as EIT, we have carried out both
   time series analysis and network analysis on this classical system.
  
The rest of this paper is organized as follows: In  Section~\ref{sec:time_series} we outline 
very briefly the salient features of time-series 
and network analysis, in 
order to make the discussion self-contained.  In  Section~\ref{sec:opto_model}, 
after introducing the quantum optomechanical model,  
we present our results on the 
time-series analysis and network characteristics in this model. The results are compared, where possible, 
with corresponding ones for the tripartite $\Lambda$ system. Section~\ref{sec:duffing_model} is devoted to a similar  study of  classical coupled Duffing oscillators. In Section~\ref{sec:conclusion}, we conclude with brief comments and indicate possible avenues for further research.

\section{\label{sec:time_series} Time-series analysis and network indicators}

\noindent  We  outline first  
the salient aspects of   time-series analysis and the manner in which an $\epsilon$-recurrence network is obtained from a  time series. The network indicators of relevance 
to us are also defined.  
Suppose we have a long time series 
${s(i)} \,  (i = 1,\,2,\cdots, M)$, 
either measured 
or otherwise generated,   
of some relevant quantity    
(the expectation value of an observable in the 
quantum mechanical case, or the value of 
a dynamical variable in the classical case).
The first task is to identify an effective phase space of dimension significantly smaller than $M$ in which the dynamics can be captured. For this purpose we need to obtain a suitable time delay $t_{d}$. Following a 
commonly  used prescription~\cite{fraser_swinney}, 
$t_{d}$ is taken to be the first minimum 
(as a function of $T$) 
of the average mutual information
\begin{equation}
I(T) = \hspace{-2ex} \sum_{s(i), s(i+T)} 
\hspace{-2.5ex} p\big(s(i), s(i+T)\big) \log_{2} \Big\{ \frac{p\big(s(i),s(i+T)\big)}{p(s(i)) p(s(i+T))} \Big\}.
\label{eqn:mutual_info_timeseries}
\end{equation}
Here,  $p(s(i))$ and $p\big(s(i+T)\big)$ are the individual probability densities for  obtaining the values $s(i)$ and $s(i+T)$ at times $i$ and $(i+ T)$, respectively, and 
$p\big(s(i), s(i+T)\big)$ is the corresponding joint probability density. Now,  employing the standard machinery of time-series analysis (see, e.g.,~\cite{abarbanel}) we 
reconstruct, from $\{s(i)\}$ and $t_{d}$, an effective phase space of dimensions $d_{\textrm{emb}}$.  In this space there are $M' = M - (d_{\textrm{emb}} - 1)t_{d}$ delay vectors  ${\bf x}_{j} \; (j = 1,\,2,\dotsc, M')$ given by
\begin{equation}
 {\mathbf x}_{j} = \big[s(j),\, s(j+t_{d}), \dotsc,\,  s\big(j + (d_{\textrm{emb}}-1)t_{d}\big)\big].
 \label{eqn:delay_vec}
\end{equation}
The underlying dynamics takes one delay (or state) 
vector to another, and phase trajectories arise, with
 $d_{\rm emb}$  Lyapunov exponents. Of direct interest to us is the maximal Lyapunov exponent (MLE),   which we have computed using the standard TISEAN package~\cite{tisean} in both the quantum and classical systems for various values of system parameters. 

  Network analysis involves coarse-graining the phase space into cells of a suitable size. An important aspect 
  here  is the construction of the adjacency matrix $A$ which depends on the cell size.  $A = R - I$, where $I$ is the $M'\times M'$ unit matrix, and  for a given cell size $\epsilon$, $R$ is the $(M'\times M')$ recurrence matrix with elements
\begin{equation}
 R_{ij} = \Theta\big(
 \epsilon -  \parallel \mathbf{x}_{i}-\mathbf{x}_{j} 
 \parallel\big).
 \label{eqn:rij}
\end{equation}
Here $\Theta$ denotes the unit step function and $\parallel \cdot\parallel$ is the standard Euclidean norm. Any 
two state vectors (equivalently, two nodes of a network) 
${\bf x}_{i}$ and ${\bf x}_{j} \,\,(i \neq j)$ are said to be 
connected iff $A_{ij} = 1$. The network is  constructed with links between such connected nodes.

The choice of the cell size $\epsilon$ is important. Its 
threshold or optimal value~$\epsilon_{c}$ must  be chosen 
judiciously.  Too  small a value of $\epsilon$ makes the network sparsely connected, with an adjacency matrix that has too many vanishing off-diagonal elements.  
Too  large a  value  of $\epsilon$ makes too many 
off-diagonal elements of $A$ equal to unity, and hence the small-scale properties of the system cannot  be captured. Our choice of $\epsilon_{c}$ is based on the recent proposal~\cite{eroglu} in the context of $\epsilon$-recurrence networks. Consider the  $(M'\times M')$ Laplacian matrix $L$ with elements 
\begin{equation}
L_{ij} = D_{ij} - A_{ij}.
 \label{eqn:lij}
\end{equation}
Here $D = {\rm diag} \,(k_{1},  \, \dotsc, \,k_{M'})$ is the degree diagonal matrix, where  $k_{i} = \sum_{j} A_{ij}$
 is the degree of node $i$. $L$ is a real symmetric matrix, and each of its row sums vanishes. Hence   the eigenvalues  of $L$ are real and non-negative, and  at least one of them  is zero. Increasing $\epsilon$ upward from zero, we determine the smallest value of $\epsilon$  (denoted by $\epsilon_{c}$)  for which the next eigenvalue of $L$ becomes nonzero.

The network indicators that we have computed for the systems of interest to us are  the average path length (APL), the link density (LD), the clustering coefficient (CC), the transitivity ($\mathcal{T}$), the assortativity ($\mathcal{R}$) and the degree 
distribution~\cite{kurths_phys_rep,boccaletti, strogatz}. 
For ready reference, their definitions are as follows. 

For a network of $M'$ nodes, the average path length APL  is given by
\begin{equation}
 \text{APL} = [M'(M'-1)]^{-1} \sum_{i,j}^{M'} d_{ij},
 \label{eqn:apl}
\end{equation}
where $d_{ij}$ is the shortest  path length connecting nodes $i$ and $j$.
The link density LD  is given by 
\begin{equation}
 \text{LD} = [M'(M'-1)]^{-1} \sum_{i}^{M'} k_{i},
 \label{eqn:ld}
\end{equation}
where $k_{i}$  is the degree of  node~$i$ (as already 
defined).
The local clustering coefficient, which measures the probability that two randomly chosen neighbors of a given node $i$ are directly connected, is defined as
\begin{equation}
 C_{i} = [k_{i}(k_{i}-1)]^{-1}  \sum_{j,k}^{M'} A_{jk} \, A_{ij}\, A_{ik}.
 \label{eqn:lcc}
\end{equation}
The global clustering coefficient CC is the arithmetic mean of the local clustering coefficients taken over all the nodes of the network.
The transitivity $\mathcal{T}$ of the network is defined as 
\begin{equation}
 \mathcal{T} =   \frac{\sum_{i,j,k}^{M'}   A_{ij}\,A_{jk}\, A_{ki}}{\sum_{i,j,k}^{M'} A_{ij} \, A_{ki}}.
 \label{eqn:transitivity}
\end{equation}
The  other indicator that we have considered is the assortativity coefficient $ \mathcal{R}$, which is  
a measure of the correlation between two nodes of a network.  Consider a randomly chosen node $j$ connected by  an edge to  a randomly chosen node $i$. Then the assortativity coefficient, also known as the Pearson correlation coefficient of degree between all such pairs of  linked nodes, is given by
\begin{equation}
 \mathcal{R} =   (1/\sigma_{q}^{2})\,\sum_{i,j}^{M'} i\,j (e_{ij} - q_{i}q_{j}), 
 \label{eqn:assortativity}
\end{equation}
where the quantities on the right-hand side are defined as 
follows. $q_{i}$ is the distribution of the `remaining' degrees, i.e., the number of edges leaving the node $j$ other than the one that connects the chosen $(i,\, j)$ 
pair. $e_{{ij}}$ is the joint probability distribution of these remaining degrees, normalized 
according to $\sum_{i,j}{e_{ij}} = 1$.  Also, 
$\sum _{j} e_{jk} = q_{k} 
= (k+1)\,p_{k+1}/\sum_{j} (j \,p_{j})$, where  $p_{k}$ is the degree distribution of the network, i.e., 
the probability that  a  randomly  chosen node  in  the  network  will  have degree $k$. Finally, 
 $\sigma_{q}^{2} = \sum_{k} k^{2} q_{k} - [\sum_{k}k q_{k}]^{2}$ is  the variance corresponding to 
 the  distribution $q_{k}$. 
 It is readily seen that  $-1\leqslant  \mathcal{R} \leqslant 1$.  $ \mathcal{R} = 1$ indicates perfect assortative mixing, $ \mathcal{R} = 0$ corresponds to non-assortative mixing, and  $ \mathcal{R} = -1$ implies complete dissortative mixing. 

\section{\label{sec:opto_model} The optomechanical model}
\noindent 
As stated in Section \ref{sec:level1}, the tripartite quantum system we examine  comprises a two-level atom  placed inside a Fabry-P\'{e}rot cavity with a vibrating mirror attached to one of the cavity walls which is capable of small oscillations. The mirror is modeled as a quantum harmonic oscillator. The model Hamiltonian  (setting $\hbar$ = 1) is given by~\cite{lahap}
\begin{align}
  H = \omega\, a^{\dagger} a & + \omega_{m}\, b^{\dagger} b + \tfrac{1}{2}\omega_{0} \sigma_{z} - G \,  a^{\dagger} a (b + b^{\dagger}) \nonumber \\
                                          &+ \Omega\, [a\, f(N) \, \sigma_{+} +  f(N) \,a^{\dagger}\,\sigma_{-}].
 \label{eqn:parent_hamiltonian}
\end{align}
$a^{\dagger}, a$ are the photon creation and annihilation operators of the cavity mode of 
frequency~$\omega$;  $b^{\dagger}, b$ are the phonon creation and annihilation operators of the mirror-oscillator unit, with  natural  frequency~$\omega_{m}$. The optomechanical coupling coefficient $G = (2\,m\,\omega_{m})^{-1/2}\, (\omega/L)$ where  $L$ and $m$ are the length of the cavity and the mass of the  mirror. The atomic operators are $\sigma_{z} = \ket{e}\bra{e} - \ket{g}\bra{g}$, \,$\sigma_{+} = \ket{e}\bra{g}$ and $\sigma_{-} = \ket{g}\bra{e}$, where   $\ket{g}$ and $\ket{e}$ denote the  ground and excited states of the atom. $\omega_{0}$ is the atomic transition frequency and $\Omega$ is the field-atom coupling constant. We have used the resonance condition $\omega = \omega_{0} + \omega_{m}$  in our analysis.  The real-valued function $f(N) = (1+ \kappa\, N)^{1/2}$ where $N = a^{\dagger}a$ is the photon number operator and $\kappa$ ($0\leqslant \kappa \leqslant 1$) is the tunable intensity parameter.  $f(N)$ incorporates 
the intensity-dependent field-atom coupling  present in the system. $N\ket{n} = n\ket{n}$ where $\ket{n}$ is the $n$-photon state.

 An effective Hamiltonian $H_{\textrm{eff}}$ for this system can be obtained~\cite{lahap}  from $H$ in the regime 
 $\omega_{m} \gg G, \,\Omega$. This is given by
\begin{align}
H_{\textrm{eff}}  = (G^{2}/\omega_{m}) &\Big\{ \beta[f(N) a^{\dagger} b \sigma_{-} + a f(N)b^{\dagger} \sigma_{+}\big]   \nonumber \\
                             &- \beta^{2} \big[a^{\dagger} a\, \sigma_{z} - \sigma_{+}\sigma_{-}\big] - (a^{\dagger} a)^{2}\Big\}. 
\label{eqn:eff_hamiltonian}
\end{align}
In real experiments the numerical values of  $G$ and $\Omega$  are comparable. Therefore, in deriving Eq. \eqref{eqn:eff_hamiltonian}, we have set $\Omega = \beta\, G$, where $\beta$ is a constant of proportionality of the order of unity.  We  investigate the dynamics of the system in terms of the dimensionless time 
$\tau = (G^{2}/\omega_{m})t$. 

The initial state  $\ket{\psi(0)}$
of the full system is taken to be a direct product of  the following states: (i)  the field in the standard normalized oscillator coherent state (CS) 
$\ket{\alpha}$, 
$\alpha\in \mathbb{C}$; 
(ii) the mirror in the oscillator ground state $\ket{0}$;  
and  (iii) the atom in an arbitrary superposition $(\cos\phi \ket{e} + \sin\phi \ket{g})$. Thus
 \begin{equation}
 \ket{\psi(0)} 
  = \sum_{n=0}^{\infty} l_{n}(\alpha) \big(\cos\phi \ket{n; 0; e} + \sin\phi \ket{n; 0; g}\big)
\label{eqn:init_state}
\end{equation}
where 
$l_{n}(\alpha) = e^{-\vert \alpha \vert^{2}/2} \alpha^{n}/\sqrt{n!}$ and the notation in the kets 
representing  product states  is self-evident.  The state of the system at any time  $t > 0$ 
  is obtained by solving the Schr\"{o}dinger equation, and is found to be given by
\begin{align}
\ket{\psi(t)} = \sum_{n=0}^{\infty}  &l_{n}(\alpha) 
\big[A_{n}(t) \ket{n; 0; e} +  B_{n}(t) \ket{n; 0; g}\big] \nonumber \\
 &+ \sum_{n=1}^{\infty} l_{n}(\alpha) C_{n}(t) \ket{n-1; 1; e}.
\label{eqn:final_state}
\end{align}
The time-dependent coefficients are given by 
\begin{subequations}
\begin{align}
 \label{eqn:soln_a}
 A_{n}(t) &= e^{i \gamma_{1} t}\cos\phi,\\
 \label{eqn:soln_b}
 B_{n}(t) &=  e^{i \gamma_{2} t} \sin\phi \, \big[\cos\,(R t) + \Delta_{b} \sin\,(R t)\big],\\ 
 \label{eqn:soln_c}
 C_{n}(t) &= e^{i \gamma_{2} t} \Delta_{c}  
 \sin\phi\,\sin(R t), 
 \end{align}
\end{subequations}
where (in units of $G^{2}/\omega_{m}$) 
\begin{subequations}
\begin{align}
 \gamma_{1} &= n^{2} + \beta^{2}(n+1), \\
 \gamma_{2} &= n^{2} - n + \tfrac{1}{2},\\
 \Delta_{b}     &= -i (n -\tfrac{1}{2}  -  \beta^{2} n)/R, \\
 \Delta_{c}     &= -i \beta\,\sqrt{n}\, f(n)/R,\\
 R  &= \big\{(n^{2} - n+\tfrac{1}{2})^{2} + \beta^{2}\, n\, f^{2}(n)  \nonumber \\
   &\qq- n[(n-1)^{2} +  \beta^{2}\, n] (n -  \beta^{2}) \big\}^{1/2}. 
 \label{eqn:parameters}
\end{align}
\end{subequations}


We now present our results. We have varied $\kappa$ from 0 to 1,  and for each value  of $\kappa$,  numerically generated a long time series  of the mean photon 
number $\aver{N}$ with time step $\delta\,\tau = 2.5\times10^{-5}$.
 After discarding the initial transients (the first $10^{4}$ points) from each of the data sets,
\begin{figure}[h]
\centering
\includegraphics[height=4cm, width=7.50cm]{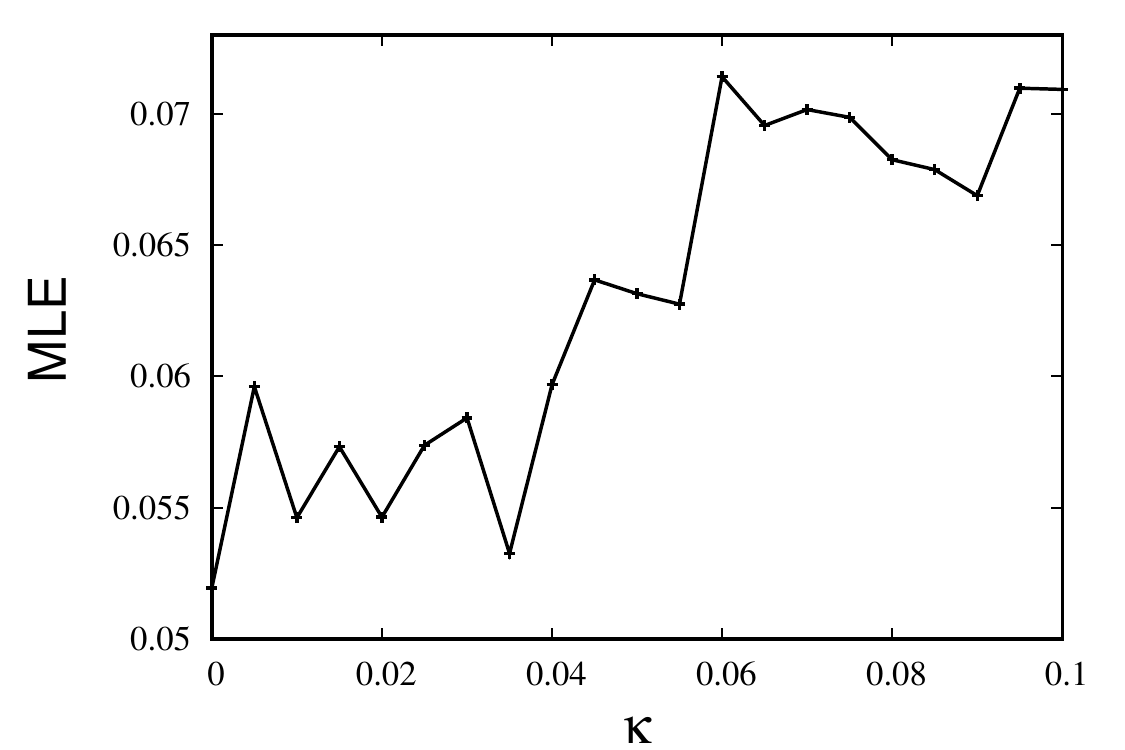}
\vspace{-1ex}
\caption{Tripartite optomechanical model: MLE versus $\kappa$ using a long time series of  $ 3\times 10^{5}$ data points.}
\label{fig:mle_300k_vs_k_opto}
\end{figure}
\begin{figure}[h]
\centering
\includegraphics[height=4cm, width=8.50cm]{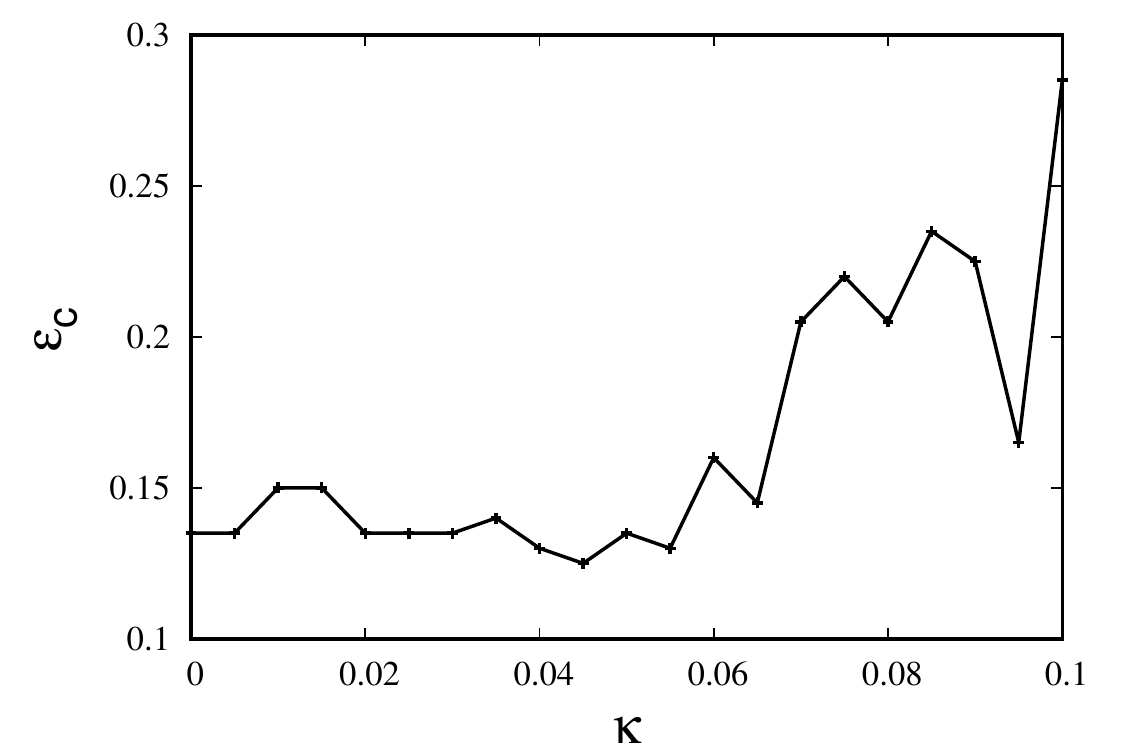}
\vspace{-1ex}
\caption{Tripartite optomechanical model: $\epsilon_{c}$ versus $\kappa$.}
\label{fig:eps_vs_k_opto}
\end{figure}
 we have  examined the manner in which  the MLE, return-time distributions, recurrence plots, etc. change  when the value of $\kappa$ is changed.  Consistent with experiments~\cite{cleland,hood}, we set $\vert\alpha\vert^{2} = 25, \,\theta = \tfrac{1}{2} \pi,\, 
\Omega = 10^{6}$\,Hz, $\beta = 1$ and $\omega_{m} =  10^{9}$\,Hz.  
Figure  \ref{fig:mle_300k_vs_k_opto} 
shows how the MLE varies with $\kappa$, 
based on a long time series of 
$3\times 10^{5}$ data points 
for each value of $\kappa$. 
We note that the reconstructed dynamics is chaotic, but only weakly so, as indicated by the small positive values of the MLE. 

We now examine how  network indicators behave as a function of $\kappa$.   In the spirit of network analysis,  for each value of $\kappa$, we have considered only 25000 data points in the corresponding time series. (This is only $\sim 8\%$ 
of the data set of the longer time series.)
The optimal value  $\epsilon_{c}$ 
has been estimated for each value of $\kappa$ (Fig. \ref{fig:eps_vs_k_opto}).  We note that the qualitative behaviour of $\epsilon_{c}$ as a function of $\kappa$ is broadly similar  to that of  the MLE in 
Fig. \ref{fig:mle_300k_vs_k_opto}. 

\begin{figure*}[ht]
  \centering
  \includegraphics[height=3.50cm, width=5.50cm,angle=0]{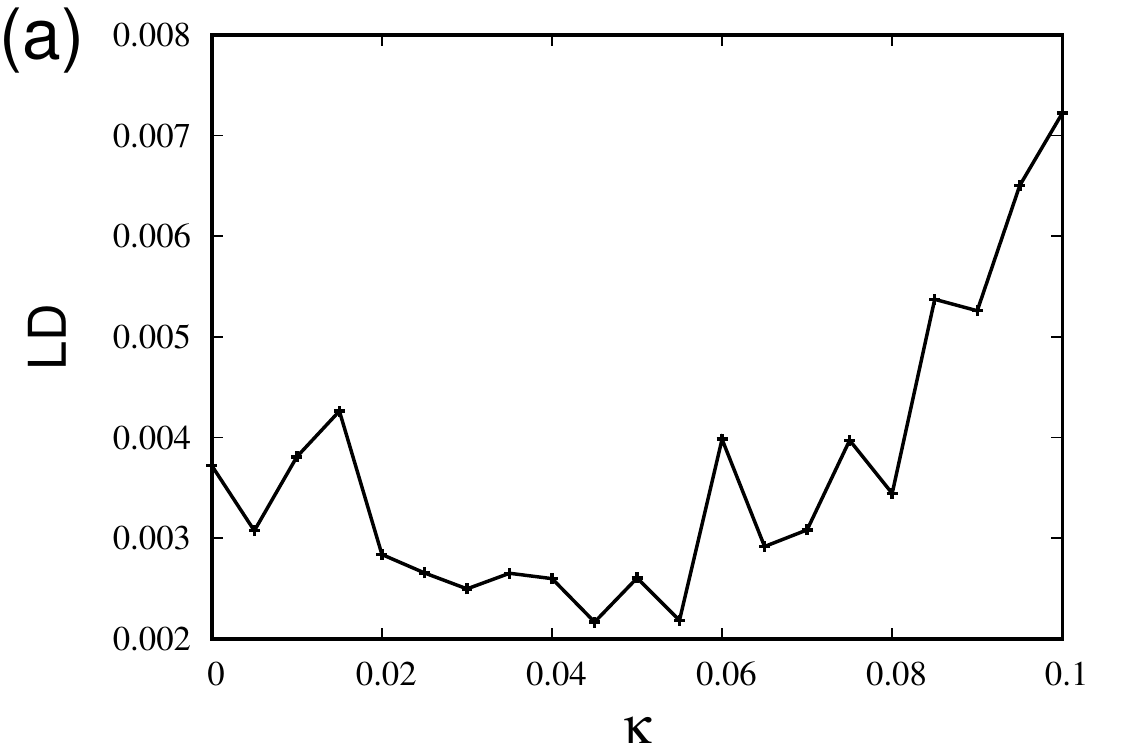}
  \includegraphics[height=3.50cm, width=5.50cm,angle=0]{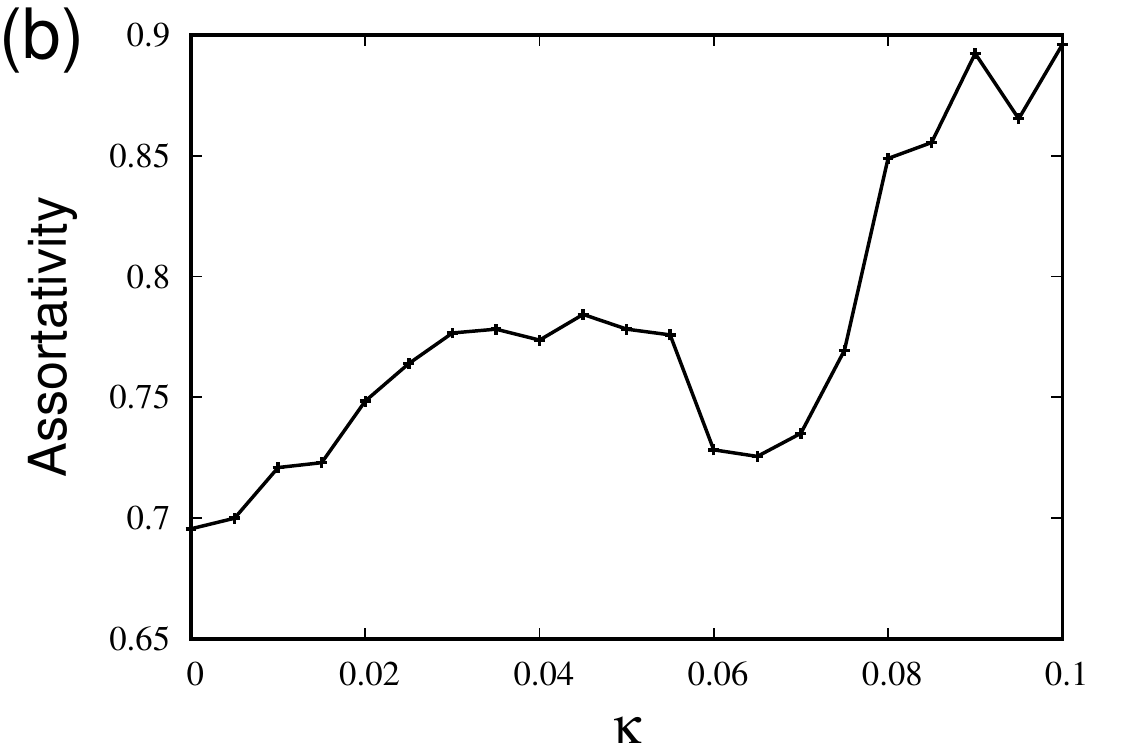}
  \includegraphics[height=3.50cm, width=5.50cm,angle=0]{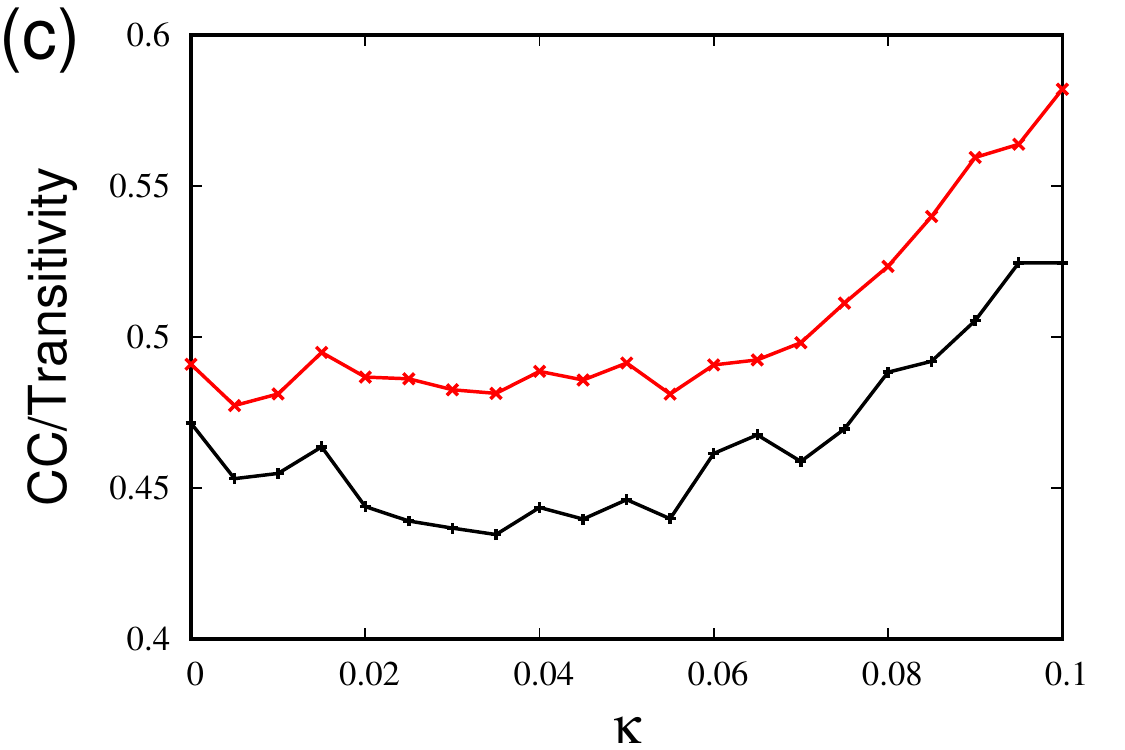}
  \vspace{-2ex}
  \caption{Tripartite optomechanical model: (a) LD, (b) assortativity,  and (c) CC (black curve)  and transitivity (red curve) versus $\kappa$. }
  \label{fig:network_opto_mech_300k_lyap}
\end{figure*}

\begin{figure}[h]
  \centering
  \includegraphics[height=3.750cm, width=8.50cm,angle=0]{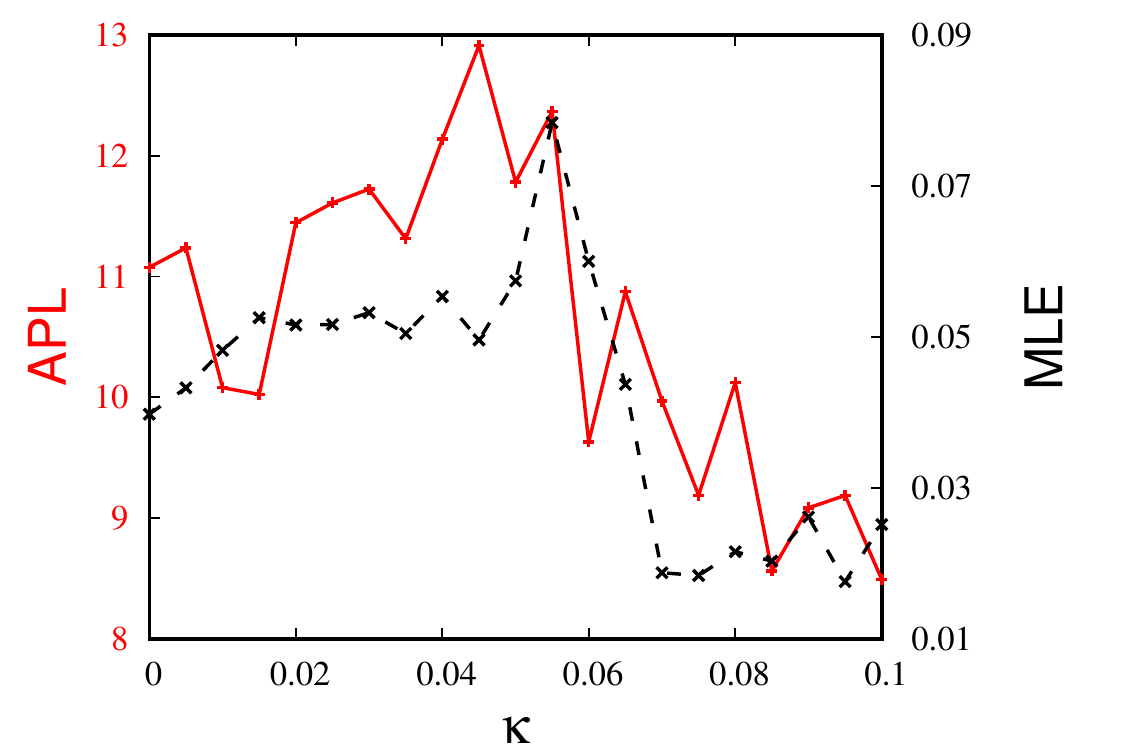}
  \vspace{-2ex}
  \caption{Tripartite optomechanical model: APL (red curve) and MLE (black dotted curve) with $25000$ data points versus $\kappa$.}
  \label{fig:network_opto_mech_25k_lyap}
 \end{figure}
\begin{figure}[h]
  \centering
  \includegraphics[height=3.50cm, width=4.250cm,angle=0]{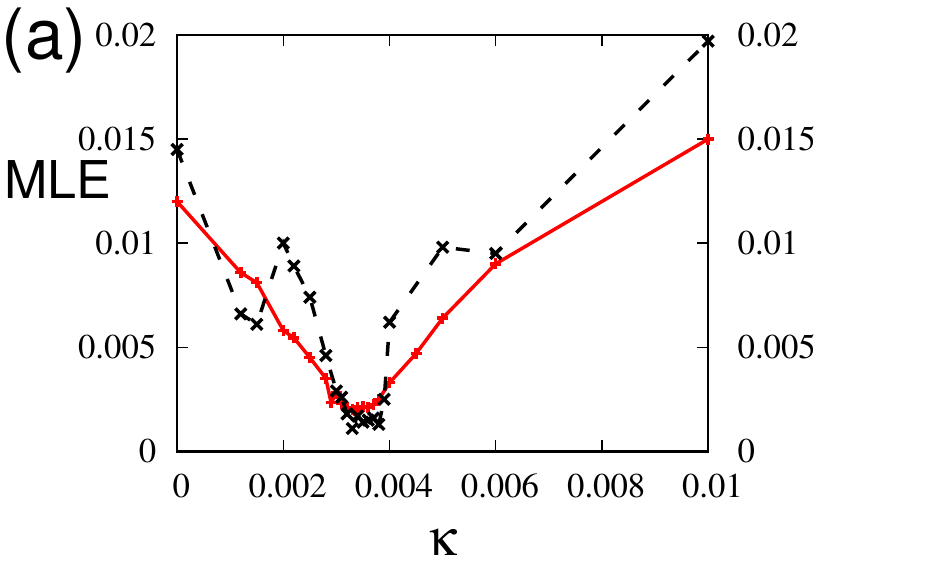}
  \includegraphics[height=3.50cm, width=4.250cm,angle=0]{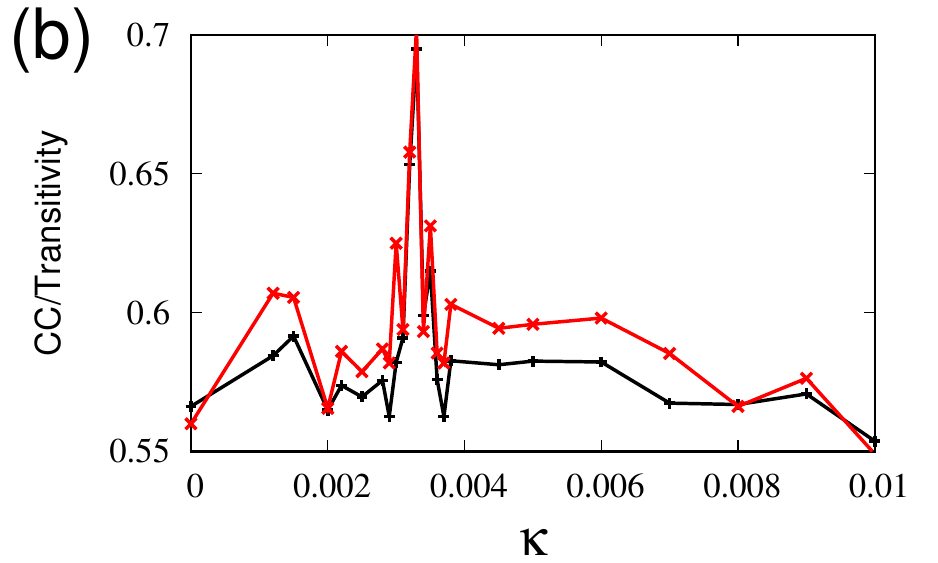}\\
  \vspace{-2ex}
  \caption{Tripartite $\Lambda$ system: (a) MLE for $3\times 10^{5}$ (black curve) and $25000$ (red curve) data points. (b)  CC (black curve)  and transitivity (red curve). These plots are repoduced from~\cite{laha4}.}
  \label{fig:lyap_cc_lambda}
 \end{figure}

 \begin{figure*}[ht]
  \centering
   \includegraphics[height=10cm, width=17.5cm,angle=0]{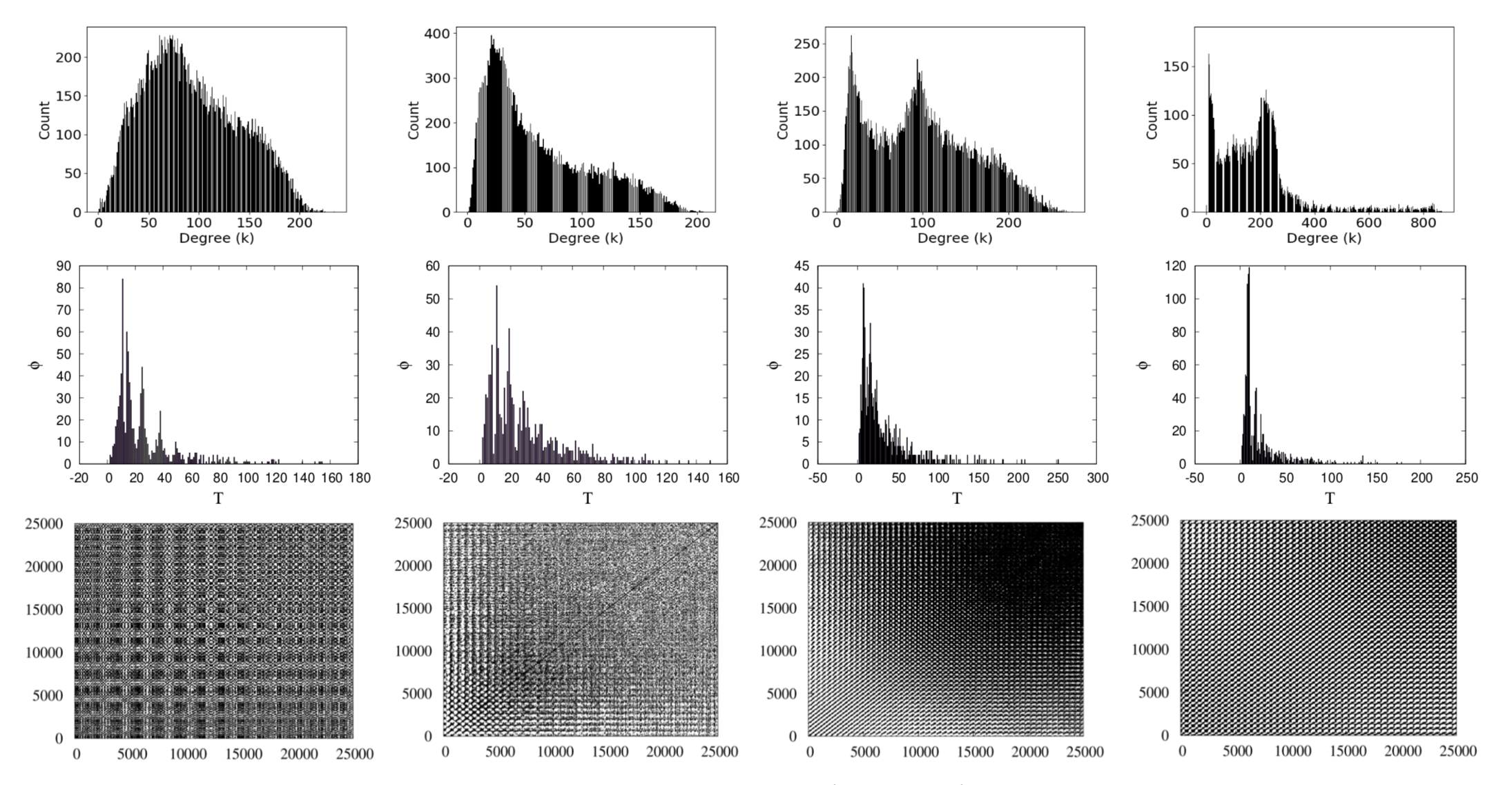}
    \caption{Tripartite optomechanical model: Degree distributions (top panel),  first-return-time distributions to 50 cells (centre panel) and recurrence plots (bottom panel) for $\kappa = $ 0, 0.03, 0.06 and 0.1 (left to right).}
\label{fig:deg_frtd_rec_plots_opto}
 \end{figure*}

The manner in which  LD, assortativity, CC and  transitivity vary with changes in $\kappa$ is  
shown in Figs. \ref{fig:network_opto_mech_300k_lyap}(a)-(c). As expected, CC and transitivity display similar behavior (Fig. \ref{fig:network_opto_mech_300k_lyap}(c)).  In all these plots,  the network indicators are obtained using the shorter time series (data sets of 25000 points). It is seen  that these network indicators  display roughly  the same  trend as $\kappa$  increases, similar to the behavior of the MLE  in Fig. \ref{fig:mle_300k_vs_k_opto}. The  APL, however, 
does not follow this trend at all (the red curve in Fig. \ref{fig:network_opto_mech_25k_lyap}). 
In this sense, LD, CC,  assortativity and transitivity are better network indicators than APL. 
On the other hand, if the MLE is 
computed using the short 
time series, its variation with $\kappa$ is 
qualitatively similar to that  of the APL
(the black dotted curve in Fig. \ref{fig:network_opto_mech_25k_lyap}), 
while differing significantly from the 
`true' variation of the MLE as depicted in 
Fig. \ref{fig:mle_300k_vs_k_opto}. 
 
These inferences  are  in  sharp contrast to those drawn from a similar  investigation  on the 
tripartite $\Lambda$ system mentioned earlier~\cite{laha4}. 
A noteworthy difference between this quantum system and the optomechanical system  is that,  for the tripartite $\Lambda$ system,  the plots of  the MLE  
versus $\kappa$  obtained with $3\times 10^{5}$ and $25000$ data points, respectively, 
  do not differ significantly (Fig. \ref{fig:lyap_cc_lambda}(a)). It has also been 
   shown in that  case that  CC and transitivity  are very good network indicators (Fig. \ref{fig:lyap_cc_lambda}(b)), and  the minimum in the MLE at $\kappa = 0.0033$ is reflected as a maximum in the CC and transitivity. 

For completeness,  we have examined the manner in which the degree distributions, return-time distributions to cells, recurrence plots, etc.  vary with changes in $\kappa$ in the optomechanical model 
under study. Each time series comprises 25000 data points. The degree distribution plots are distinctly different for different values of $\kappa$ (the top panel of Fig. \ref{fig:deg_frtd_rec_plots_opto}). For instance, the single-peaked distributions for smaller values of $\kappa$ gradually change to double-peaked distributions as $\kappa$ is increased. Further,  the spread in the distributions changes dramatically with increasing  $\kappa$. 

The first-return-time distributions to a specific cell for various values of $\kappa$ are shown in the centre panel of  Fig. \ref{fig:deg_frtd_rec_plots_opto}. We find that there exist several significant peaks apart from a prominent peak for almost all values of $\kappa$. For higher value of $\kappa$ ($> 0.06$) the spread in the distribution is relatively smaller. We have verified that the second-return-time distributions exhibit similar behavior.

The manner in which recurrence plots change with 
$\kappa$ is displayed in the bottom panel of Fig. \ref{fig:deg_frtd_rec_plots_opto}.  The return maps and the power spectra are not very sensitive to changes in $\kappa$. 
  In the tripartite $\Lambda$ system,  qualitative changes in the recurrence plots, return maps  and recurrence-time distributions  with changes in the value of 
$\kappa$ mirrored the fact that 
  the MLE was at a minimum at $\kappa = 0.0033$. Such clear signatures are  absent in the case of the optomechanical model. 

We turn in the next Section to  a classical system which is a near analog of the tripartite $\Lambda$ system, namely, two coupled Duffing oscillators. In this case, 
it is shown that the variation of the MLE with changes in a parameter analogous to 
$\kappa$ are similar  for data sets with  $10^{5}$ points and  $25000$ points respectively, in the sense that both show an  overall increase with as 
$\kappa$ is increased (Fig. \ref{fig:lyap_expo_duffing}). This feature is akin to that displayed in Fig. \ref{fig:lyap_cc_lambda}(a) for the tripartite 
$\Lambda$ system, although in that  case the sensitivity to the number of data points is significantly lower. It is therefore worth investigating the differences in the behavior of network indicators in these two models.

\section{\label{sec:duffing_model} Coupled Duffing oscillators}
\noindent As mentioned 
in the Introduction, a  
 classical system comprising  two coupled oscillators driven by a harmonic force mimics~\cite{alzar} the phenomenon of EIT manifested in the tripartite quantum system comprising a $\Lambda$-atom interacting with two radiation fields. The  dynamical equations for the displacements $x_{1}$ and $x_{2}$ of the two oscillators  are given by
\begin{align}
 \ddot{x}_{1} + \delta_{1} \dot{x}_{1} + \omega_{\text{cl}}^{2} \,x_{1} - \Omega_{\text{cl}}^{2} \,x_{2} &= f \sin\,(\Omega_{d} t), \\
 \ddot{x}_{2} + \delta_{2} \dot{x}_{2} + \omega_{\text{cl}}^{2} x_{2} - \Omega_{\text{cl}}^{2} x_{1} &= 0.
\end{align}
Here, $\delta_{1}$ and  $\delta_{2}$   are damping parameters, $\omega_{\text{cl}}$ is the stiffness parameter, $\Omega_{\text{cl}}$ is the coupling parameter, and $f$ and $\Omega_{d}$ are, respectively, the amplitude and angular frequency of the periodic driving force.
\begin{figure}[h]
\centering
\includegraphics[height=4cm, width=8.50cm]{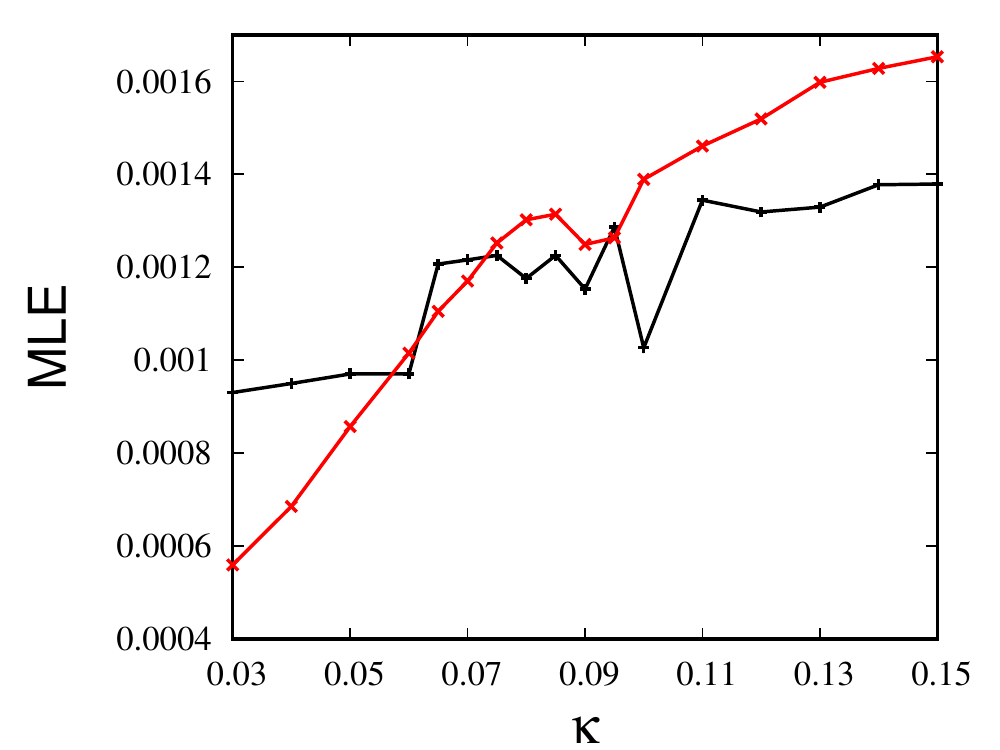}
\caption{Classical model: MLE versus $\zeta$ for  $25000$ (red curve) and $10^5$ (black curve) data points.}
\label{fig:lyap_expo_duffing}
\end{figure}
\begin{figure*}[ht]
  \centering
  \includegraphics[height=3.750cm, width=7.50cm,angle=0]{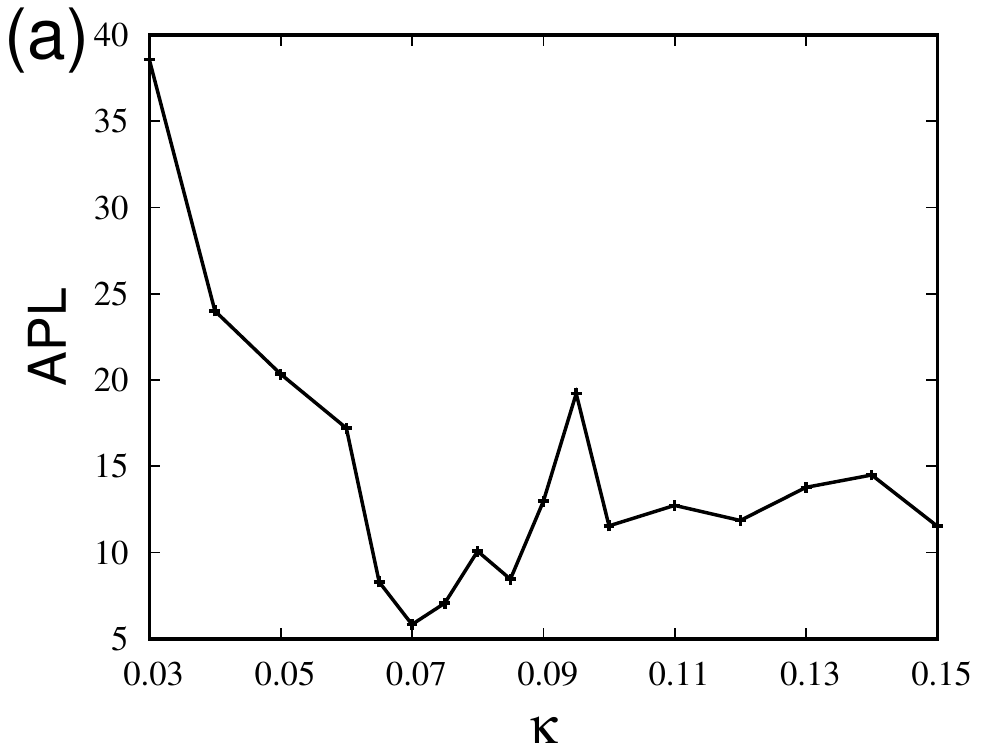}
  \includegraphics[height=3.750cm, width=7.50cm,angle=0]{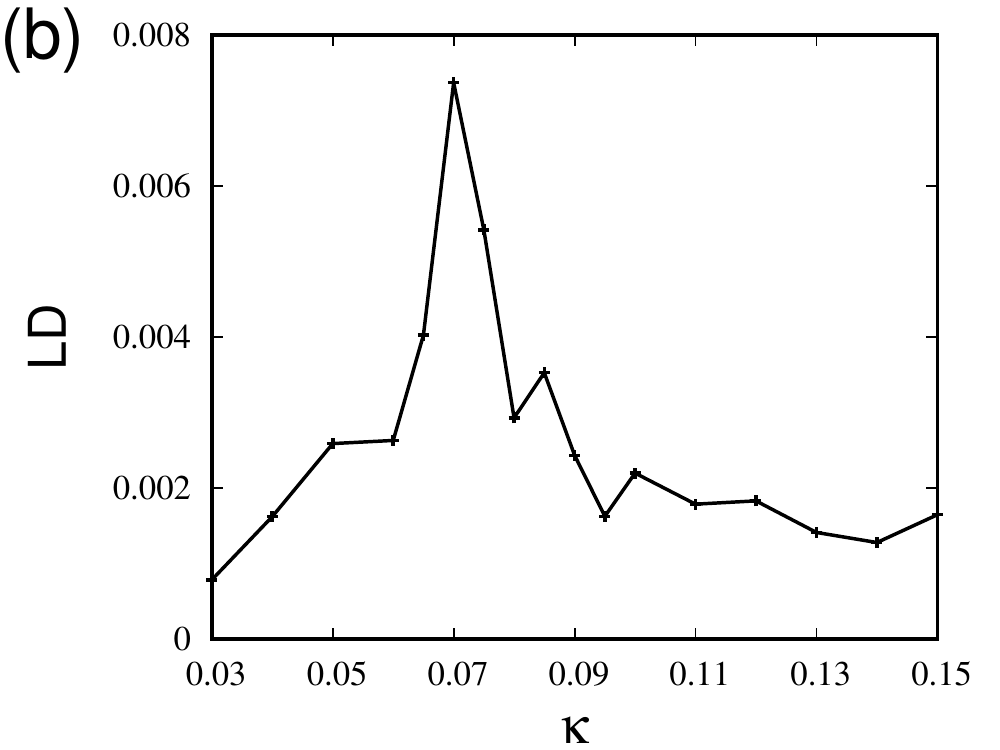}
  \includegraphics[height=3.750cm, width=7.50cm,angle=0]{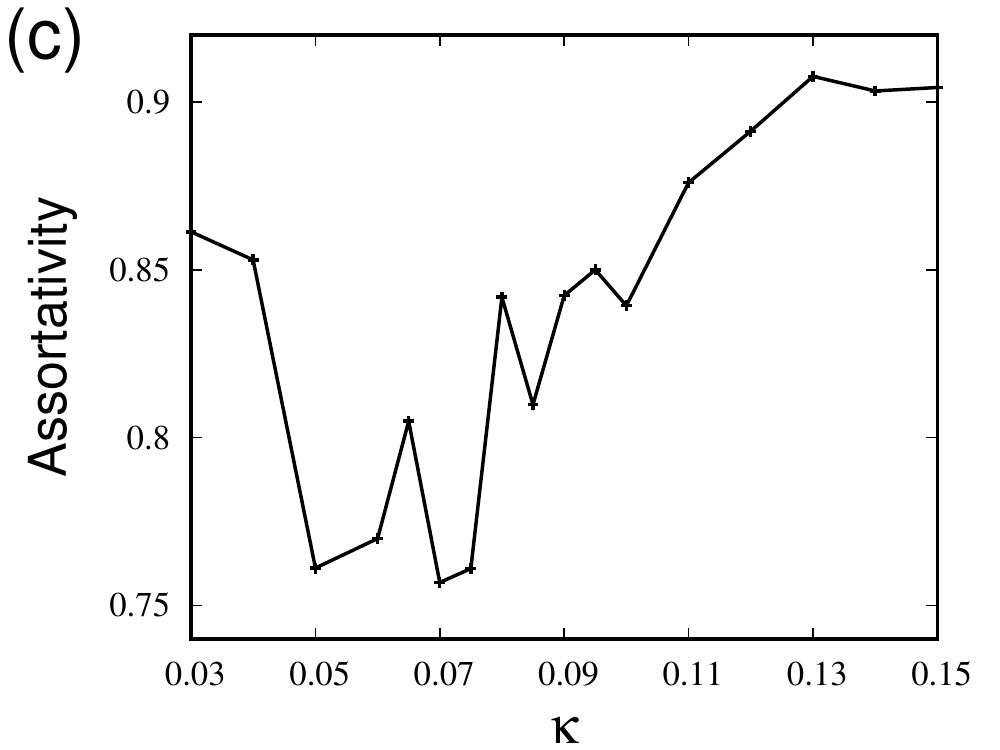}
  \includegraphics[height=3.750cm, width=7.50cm,angle=0]{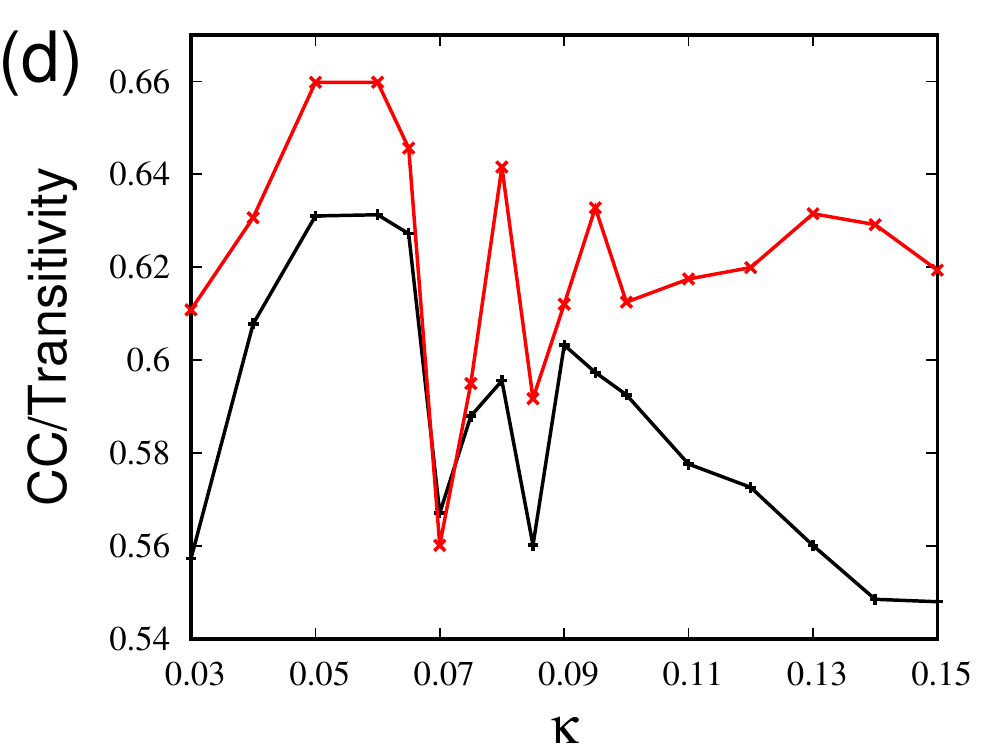}
  \caption{Classical model:  (a) APL, (b) LD, (c) assortativity,  and (d) CC (black curve)  and transitivity (red curve) versus $\kappa$.}
  \label{fig:network_duffing}
 \end{figure*}
In the quantum mechanical 
counterpart of this system,  $\delta_1$ is the strength of spontaneous emission from the excited state of the $\Lambda$-atom, 
$\delta_2$ is the energy dissipation rate of the pumping transition, and $f$ is the amplitude of the driving field.  In practice, of course, 
nonlinear effects arise. We have 
therefore considered the modified  coupled Duffing equations  
\begin{align}
\ddot{x}_{1} + \delta_{1} \dot{x}_{1} + \omega_{\text{cl}}^{2} \,x_{1} + 
\zeta \,x_{1}^{3} - \Omega_{\text{cl}}^{2}\, x_{2} &= f \sin\,(\Omega_{d} t), \\
\ddot{x}_{2} + \delta_{2} \dot{x}_{2} + \omega_{\text{cl}}^{2} \,x_{2} + \zeta \,x_{2}^{3} - \Omega_{\text{cl}}^{2} \,x_{1} &= 0, 
\end{align}
where $\zeta$, the strength of the nonlinearity, is analogous to the IDC parameter $\kappa$ in the quantum model. As is well known, the 
forced Duffing oscillator exhibits  a very 
diverse range of complex dynamical behavior, 
depending on the values of the parameters. 
 For numerical computations we have set
 the parameters at representative values 
 $\omega_{0} = 2, \,\omega_{cl} = \sqrt{10},\,
  \Omega_{cl} = \sqrt{6},\, \delta_1 = 
  10^{-2}$ and $\delta_2$ = $10^{-7}$. The dynamical variable considered is the velocity  
  $\dot{x}_{2}$. A long time series of 
  $\dot {x}_{2}$  was obtained for various values of $\zeta$ with initial conditions $x_1(0) = 1, \,\dot{x}_{1}(0) = 0, \,x_{2}(0) = 0,\, 
  \dot{x}_{2}(0) = 0$.

The manner in which the  MLE  varies 
with  $\zeta$ for  $10^{5}$ and  $25000$ data points respectively (the black curve and the red curve in Fig. \ref{fig:lyap_expo_duffing}) reveals that the gross features in the two plots are in reasonable agreement with each other, in contrast to the case of the optomechanical model.

The changes in the behavior of the APL, LD, assortativity, CC and  transitivity  with 
varying  in $\zeta$ are shown in Figs. \ref{fig:network_duffing}(a)-(d). It is interesting to note that none of these indicators seem to carry any signatures of the behavior of the MLE with $\zeta$. This is in marked contrast to the situation in both the quantum models considered above.


\section{\label{sec:conclusion} Concluding remarks}

In this work, we have carried out detailed time-series analysis and network analysis on a fully  quantum optomechanical model and on a classical model of two interacting Duffing oscillators. Nonlinearities are inherent  in 
both cases: in  the former,  the intensity-dependent coupling between subsystems; 
in the latter, a  cubic nonlinearity in each oscillator. An  archetypal indicator  of complex dynamics,  the maximal Lyapunov exponent 
(MLE), has been obtained from the analysis of 
a long time series in each case. Network analysis of the same systems has been carried out, and  network indicators have been estimated from a considerably abbreviated time series. The variations of 
these quantities with changes in the nonlinearity parameters have  been examined extensively and compared with each other.  The conclusions drawn on the similarities between these two sets of indicators are also compared with those obtained
from  earlier  investigations on a reference system (another tripartite quantum model comprising a $\Lambda$-atom interacting with two radiation fields).   

A noteworthy  feature that emerges is the following.  Network indicators such as the clustering coefficient (CC) and the transitivity capture the behavior of 
the maximal Lyapunov exponent (MLE) 
very closely  in the quantum system, provided
the latter is not very sensitive to the number of data points used,  as in the case of the reference system.  
In the optomechanical model, on the other hand,  the 
 MLE is found to be very sensitive to the size of the data set. In this instance,  
 the CC and the transitivity  merely capture the overall trend in the MLE reasonably well,  without closely following its variation with the nonlinearity parameter. In the classical model considered, while the MLE is not very sensitive to the number of data points used in the analysis, the nonlinearity does not appear in the interaction between subsystems, but only in the individual subsystems. This 
 is the likely  reason why none of the network indicators considered displays signatures of the manner in which the MLE changes with the nonlinearity.  An  important extension of this work would be the identification of  
 other `good' network indicators which reflect in 
 {\em all} cases the variations in the MLE when the  nonlinearity  is tuned,  and their sensitivity to the precise form of the  nonlinearity. 
 Network analysis would then provide a reliable  
 shorter technique than time-series analysis for determining the salient features 
of complex dynamical behavior in the expectation 
values of observables in multipartite quantum mechanical systems.

\acknowledgements  
We acknowledge Soumyabrata Paul for help with some numerical computations in the classical model.

\bibliography{reference}

\end{document}